
\magnification=1200
\centerline{NICOLAUS COPERNICUS ASTRONOMICAL CENTER, Warsaw, Poland}
\vskip0.2in
\centerline{~~~~~~~~~~~~~~~~~~~~~~~~~~~~~~~~~~~~~~~~~~~~~~~~~~~~~~~~~No. 291}
\centerline{~~~~~~~~~~~~~~~~~~~~~~~~~~~~~~~~~~~~~~~~~~~~~~~~~~~ January 1995}
\centerline{~~~~~~~~~~~~~~~~~~~~~~~~~~~~~~~~~~~~~~~~~~~~~~~~~~~ hep-ph/9502310}
\vskip0.6in
\centerline{\bf Equation of State and Temperature of First Heavy Particles}
\centerline{\bf Arising in the Universe at the Grand Unification Scale}
\vskip0.2in
\centerline{ I. G. Dymnikova}
\vskip0.1in
\centerline{\sl Nicolaus Copernicus Astronomical Center}
\centerline{\sl Bartycka 18, 00-716 Warsaw, Poland}
\vskip0.1in
\centerline{\sl N. Copernicus Foundation for the Polish Astronomy}
\centerline{\sl Al. Ujazdowskie 4, 00-478 Warsaw, Poland}
\vskip0.2in
\centerline{ M. Krawczyk}
\vskip0.1in
\centerline{\sl Institute of Theoretical Physics, Warsaw University}
\centerline{\sl Hoza 69, PL-00-681 Warsaw, Poland}
\vskip0.4in
\centerline{Submitted to "Astrophysics and Space Science"}

\vfill\eject

\centerline{\bf EQUATION OF STATE AND TEMPERATURE OF FIRST HEAVY PARTICLES}
\centerline{\bf ARISING IN THE UNIVERSE AT THE GRAND UNIFICATION SCALE}
\vskip0.3in
\centerline{ I. Dymnikova}
\vskip0.1in
\centerline{\sl N. Copernicus Astronomical Center}
\centerline{\sl Bartycka 18, PL-00-716 Warsaw, Poland}
\vskip0.1in
\centerline{\sl N. Copernicus Foundation for Polish Astronomy}
\centerline{\sl Al. Ujazdowskie 4, PL-00-478 Warsaw, Poland}
\vskip0.2in
\centerline{ M. Krawczyk}
\vskip0.1in
\centerline{\sl Institute of Theoretical Physics, Warsaw University}
\centerline{\sl Hoza 69, PL-00-681 Warsaw, Poland}
\vskip0.2in
\centerline{\bf Abstract}
We consider
first heavy particles with masses
$M\sim M_{GUT}$ which arise in the Universe during phase transitions
at the Grand Unification scale. Using statistical mechanics approach
we show that they
behave like ideal quantum degenerate Bose gas
which has the temperature of the Hawking thermal
radiation due to presence of the de Sitter
event horizon.
The equation of state for both scalar and gauge bosons is presented
including the coupling constant at the Grand Unification scale.
\vskip0.4in

According to Grand Unified theories,
first phase transitions occur at the Grand Unification scale
$E_{GUT}\sim{(10^{14}-10^{16}) GeV}$, and the first particles which
are scalar and gauge bosons get masses of order $M\sim M_{GUT}$ via
spontaneous symmetry breaking (SSB).
In the context of inflationary models related to
a symmetry-breaking phase transition, scalar field
reaches the minimum of its effective potential at the end of inflation.
A vacuum energy then exists in the form of coherent oscillations
of a scalar field around its minimum, corresponding to a condensate
of zero-momentum $\phi$ particles
with masses $M\sim M_{GUT}$ ${^{1}}$.
In models with chaotic inflation coherent oscillations
of scalar field are described by a dust equation of state
corresponding to massive particles at rest ${^{2}}$.

In the present paper we consider first heavy nonrelativistic
particles using statistical mechanics as the model-free tool.
Such an approach allows to reveal characteristic features of their behaviour
which would not depend crucially on details of particular inflationary
scenarios.

The first phase transition can be related to the global discrete SSB, because
the arrow of time is expected to appear then ${^{3}}$.
Hence scalar bosons can arise first in the Universe.
Heavy nonrelativistic gauge bosons are expected
to emerge next via the local SSB.

These particles seem to be really first particles in the Universe,
in accordance with
the Turner "No-Hair Theorem": "Inflation greatly lessens dependence of the
present state of the Universe upon its initial state"${^{2}}$.

This statement goes back to the Sakharov's ${^{4}}$ and Gliner's ${^{5}}$
suggestion, formulated in 1965, that the vacuum state
$$p = -\varepsilon\eqno(1)$$
can be the limiting state at superhigh densities, and to the Gliner's
1970 suggestion ${^{6}}$
that it can be an initial state for the expanding Universe.
The cosmological model describing evolution of the Universe
starting from the initial vacuum state (1) has been calculated
in 1975. Using simple phenomenological equation of state
for transition from an inflationary stage (1) to radiation-dominated
stage $p=\varepsilon/3$, it has been shown that the de Sitter vacuum state (1)
provides both reason for expansion and isotropy and homogeneity of the
Universe, independently on mechanisms responsible for its arising ${^{7}}$.

In 1981 Guth has found that the equation of state (1) can arise
in Grand Unified Theories
as the equation of state for a scalar field in a metastable state called
the false vacuum ${^{8}}$.
Since 1981 most attention was focused on constructing
physical mechanisms driving inflation and providing further reheating
of the Universe (for current status of reheating problem
see ${^{9}}$ and references there).

On the other hand, in  the context of SSB, the very first postinflationary
material can be made up from heavy nonrelativistic particles with masses
$M\sim M_{GUT}$.

Our aim is to asses its characteristic properties which would be connected
with main physical features of inflation and would not depend on details
of particular models.

The energy scale of first SSB phase transitions is $E\sim E_{GUT}$.
At the beginning and during phase transitions the Universe is
expanding in the superluminal way ${^{1}}$.
Therefore first heavy particles
emerge as test particles in the vacuum (1)
with the density  $$\rho_{vac}\sim\rho_{GUT}=\rho_{Planck}
\biggl({{E_{GUT}}\over{E_{Planck}}}\biggr)^4\eqno(2)$$
They can interact within the causally connected region
confined by the de Sitter event horizon
$$a_{0} = \biggl({{3c^2}\over{8\pi G\rho_{vac}}}\biggr)^{1/2}
=\biggl({{3\over{8\pi}}\biggr)^{1/2}}\biggl({{E_{Pl}}\over{E_{GUT}}}
\biggr)^2 l_{Pl}\eqno(3)$$
This can be the reason why {\it heavy} particles arise first. Their
Compton wavelength
$${\lambda}\sim{{\hbar}\over{M c}}\sim{\biggl({{E_{Pl}}\over{E_{GUT}}}
\biggr)l_{Pl}}\eqno(4)$$
can not exceed radius of causally connected region (3).

The lower limit for the particles kinetic energy follows from the uncertainty
relation as
$$E_{min}\sim{{p^{2}}\over{M_{GUT}}}
\sim{{8\pi}\over 3}\biggl({{E_{GUT}}\over{E_{Pl}}}\biggr)^2 E_{GUT},\eqno(5)$$
where $p\sim {{\hbar}/{a_{0}}}$.
For $E_{GUT}\sim{10^{15}GeV}$ we have
$$\rho_{GUT}\sim{10^{77}g/cm^3};~~~~~
a_0\sim{10^{-25}cm};~~~~~\lambda\sim{10^{-29}cm};~~~~~
E_{min}\sim {{10^7} GeV}\eqno(6)$$
We can estimate the upper limit
for $E_{kin}$ taking into account that
newborn particles will immediately "see"
the Hawking thermal
radiation due to presence of the de Sitter event horizon.
Its temperature is determined by the Hawking formula ${^{10}}$
$$kT_{H} = {1\over{2\pi}}{\biggl({{l_{Pl}}\over{a_{0}}}\biggr)}E_{Pl}
= \biggl({2\over{3\pi}}\biggr)^{1/2}
{\biggl({{E_{GUT}}\over{E_{Pl}}}\biggr)}E_{GUT}
\sim{10^{-4}E_{GUT}}\sim10^{11}GeV,\eqno(7)$$
where $k$ is the Boltzmann constant.

Since the upper limit for kinetic energy $E_{max}\sim{kT_H}\ll E_{GUT}$,
one can not expect
that particles have large momenta. Therefore
the CM energy of
their collisions $\sqrt s $
is close to the sum of rest energies of the colliding particles
$\sqrt s \sim 2 E $.
As a result the dominating
interactions are due to elastic 2 $\rightarrow$ 2 body scatterings.

For the first scalar particles, which are selfinteracting,
we can estimate characteristic radius of interaction.
Since particles mediating interaction are the same
scalars with masses $M_{sc}$,
the cross section
depending on  the coupling constant $g_{sc}$
is determined by
$$\sigma_{int}\sim {{{\alpha_{sc}}^2}\over{M_{sc}^2}}
\sim {\alpha_{sc}}^2 \lambda^2,
\eqno(8)$$
where $$ \alpha_{sc}=g_{sc}^2/4\pi$$
The characteristic radius of interaction (defined by the relation
$\sigma_{int}\sim{ r_{int}^2}$)
$$r_{int}\sim{\alpha_{sc}\lambda}\eqno(9)$$
is smaller than $\lambda$,
if $\alpha_{sc}$ is smaller than unity.
For heavy gauge bosons (X,Y) emerging via the local SSB,
with
masses $M_{x}\sim M_{GUT}$, the coupling constant
$\alpha_x$ is
estimated as ${^{11,12}}$
$$\alpha_x = g{_x}^2/{4\pi}\sim{1/40 - 1/25}\eqno(10)$$
Thus, for heavy gauge bosons
$r_{int}<\lambda$  due to (10).
It is rather natural to expect that $\alpha_{sc}\sim \alpha_x$
(in SUSY case $g_{sc}=g_x$),
then the
cross-section (8) is small enough for particle-particle interaction to
produce a temperature exceeding (7).

With the upper limit on kinetic energy given by $(7)$, we obtain the ratio
$${{E_{kin}}\over{E_{GUT}}}\leq
{{E_{max}}\over{E_{GUT}}}\sim{\biggl({2\over{3\pi}}\biggr)^{1/2}
\biggl({{E_{GUT}}\over{E_{Pl}}}\biggr)}
\sim {10^{-4}}\eqno(11)$$
This quantity
characterizes the accuracy with which we can consider first heavy
particles as a dust with the equation of state $p = 0$.

In order to proceed further
we should make sure that we have ensemble of particles
to apply statistics. This is just the moment when we need cosmological model
of transition.

There is no generally accepted detailed theory now,
describing a transition from the inflationary
stage of the Universe evolution to a postinflationary stage (see, e.g.,
${^{9}}$). To make our estimates as model-free as possible, we describe
a transition from
the vacuum state (1) to the dust stage, using the simple phenomenological
equation of state which
is not related to a particular inflationary model and
contains least possible number of arbitrary parameters (it has been used
in 1975 to connect the inflationary onset of the Universe
with the standard scenario of its evolution afterwards).
It has the form
${^{7,13}}$
$$p + \varepsilon = {\varepsilon_{1}}{{{(\varepsilon_{0} - \varepsilon)}^
{\beta}}\over
{{(\varepsilon_{0} - \varepsilon_{1})}^{\beta}}}\eqno(12)$$
At $\varepsilon = \varepsilon_{0}$ it gives the vacuum equation of state $(1)$.
At $\varepsilon = \varepsilon_{1}$ it gives the dust equation of state $p = 0$.
A parameter $0<\beta\leq 1$ characterizes the rate of a transition.

Integrating the Friedmann equations with the transitional equation
$(12)$, we can describe evolution of a scale factor and energy density
in the analytic form in the cases $\beta\rightarrow 1$,
$\beta\ll 1$ and $\beta=1/2$.
For $\beta=1/2$, we obtain the scale factor
$$ a=a_{0}{\exp{\biggl({A\sin{{ct}\over{Aa_{0}}}}\biggr)}},\eqno(13a)$$
where $A=
2{{\sqrt{(1 - \varepsilon_{1}/\varepsilon_{0})}}/
{(3\varepsilon_{1}/\varepsilon_{0})}}$.
Evolution of energy density during a transition is given by
$$\varepsilon=\varepsilon_{0}\biggl(1 - \sin^2{{ct}\over{Aa_{0}}}\biggr)
\eqno(13b)$$
In the case $\beta\ll 1$ the scale factor evolves as
$$a=a_0\exp\biggl({1\over2}(1+\sqrt{\varepsilon/\varepsilon_o}){{ct}\over{a_0}}
\biggr)\eqno(14a)$$
and energy density as
$$\varepsilon=\varepsilon_0\biggl(3{\varepsilon_1\over\varepsilon_0}
{{ct}\over{a_0}}-1\biggr)^2\eqno(14b)$$

To estimate the duration of transition $t_1$,
we use the assumption that
the whole (or almost whole) rest mass has been created during phase
transitions at the Grand Unification scale in the form of heavy particles
with masses $M\sim M_{GUT}$, which later have decayed into some more light
species, but no (or almost no) new rest mass has been created during reheating
and during further expansion of the Universe.
This assumption looks natural because
in the course of reheating ultrarelativistic particles can be created
whose total energy in any case can not exceed the characteristic energy
$E\sim E_{GUT}$ and whose rest mass energy would be therefore much
smaller than this quantity.

Applying condition of conservation of the rest mass
$a_{1}^3\rho_{1}=a_{today}^3\rho_{today}$,
we obtain
$$\rho_{1}\sim(1/65)\rho_{GUT};~~~~~
a_{1}\sim10^{-6}cm;~~~~~
t_{1}\sim 2\times10^{-34}s\eqno(15)$$
for $\rho_{today}\sim10^{-29}g/cm^{3}$.
These estimates do not depend crucially on the choice of the value of the
parameter $\beta$ in the equation (12). The most noticeable difference
takes place for the parameter $\rho_1$ which at $\beta\ll 1$
takes the value $\rho_1\sim{(1/120)\rho_{GUT}}$. (The case
$\beta\rightarrow1$ corresponds to the power-law inflation.)

The transition into the dust implies that the whole rest mass
($M_{1}\sim 2\times 10^{57}g$) is contained in particles with mass
$M_{GUT}\sim{2\times 10^{-9}g}$.
The number of particles is $N_{1}\sim 10^{66}$.
The volume at the end of transition is $V_{1}\sim 4\times 10^{-18}cm^3$,
hence density of particles is
$$n={{N_{1}}\over{V_{1}}}\sim{10^{84}cm^{-3}}\eqno(16)$$

Using the Friedmann equations and formulae (13)-(14), we can check total
energy balance and find that only about $1/4$ of
vacuum energy involved in transition has been spent. The substantial amount
remains available for further expansion and reheating. Therefore the Universe
is still mainly vacuum dominated with $\rho_{vac}\sim{0.7\rho_{GUT}}$,
and particles can interact within
the causally connected regions
confined by the de Sitter horizon
$(3)$. The number of particles in
such a region is
$N\sim10^{10}$.
Hence, we have an ensemble and can use statistics.

According to (16), the mean distance between particles is
$$r_{av}\sim 10^{-28}cm\eqno(17)$$
The characteristic radius of interaction $(9)$
is much smaller than $r_{av}$ and smaller than $\lambda$
(for $\alpha_s < 1$ and $\alpha_x < 1$).
Therefore the free path $l$
is determined by geometrical cross section
$\sigma_0\sim{\pi{\lambda}^2}$
(in hard balls approximation
with $\lambda$ as characteristic size of particles) as
$$l \sim{1\over{n\pi\lambda^2}}\sim 3\times 10^{-27}cm\eqno(18)$$
As a result our particles satisfy two general criteria for
ideal gas:
$$n r^{3}_{int}\ll 1 ~~~~~~~~~~~    r_{int}\ll l$$
On the other hand, $l\ll a_0$ which means that in causally connected
regions particles can collide, and their equation of state would differ
from dust equation of state $p=0$.

Using the upper limit for kinetic energy put by (7),
we can estimate the lower limit for the de Broglie wavelength as
$$\lambda_{de Broglie}\sim{{\hbar}\over{{{p}}_{H}}}\sim{10^{-27}}cm
\eqno(19)$$
Since $\lambda_{de Broglie} > r_{av}$, our ideal gas is essentially quantum
one. Number of particles is much higher
than the number of quantum states they can populate:
$${\biggl({{\lambda_{de Broglie}}\over{r_{av}}}\biggr)^3}=
{{N_1}\over{V_1}}\cdot
{{\hbar^3}\over{p^3}}={{N_1}\over{(V_1 p^3)/\hbar^3}}\sim
n{{\hbar^3}\over{(M kT_H)^{3/2}}}\gg1\eqno(20)$$
The temperature of degeneration for ideal quantum Bose gas
is determined by ${^{14}}$
$$k T_{deg}= 3.31 {{\hbar^2}\over{g_{st}^{2/3}M}} n^{2/3}\eqno(21)$$
Here $g_{st}=2s+1$ is statistical weight, with $s=0$ for scalar and
$s=1$ for vector bosons.

For our gas $k T_{deg}\sim10^{14}GeV$.
Because it interacts with the Hawking
thermal radiation its temperature could be close to $T_{H}$.
Let us show that this is the case.
Interaction of thermal radiation $(7)$ with
nonrelativistic heavy small particles would be essentially the Thomson
scattering, because the condition for low-freqency limit
$${{E_{kin}}\over{M c^2}}{{\hbar\nu}\over{M c^2}}\sim
{\biggl({{kT_H}\over{E}}\biggr)}^2\ll1$$
is satisfied.
In our case the Thomson cross section can be written as
$${\sigma}_{T} = {{8\pi}\over 3}{{{\alpha}^{2}{\hbar}^{2}}\over{M^{2}c^2}}=
{{8\pi}\over 3}{\alpha^2}{\lambda}^2\eqno(22)$$
with $\alpha$ for both $\alpha_{sc}$ and $\alpha_x$.
The time for achieving thermal equilibrium between particles and radiation
is the relaxation time
$$\tau_{rel}^{p\gamma} = {{1\over{n{\sigma}_{T} c}}}\sim
{5\times 10^{-35}s}\eqno(23)$$
Because $\tau_{rel}^{p\gamma}<{t_1}$, thermal equilibrium can be
achieved during a transition, and
particles can acquire
the Hawking temperature $T_H$.
Since
${T_{H}}\ll{T_{deg}}$, we have {\it ideal quantum degenerate
Bose gas} with the equation of state ${^{14}}$
 $$P = 0.0851 g_{st}{{M^{3/2}}\over{\hbar^3}}
(kT_H)^{5/2}\eqno(24)$$
We see that the process of emerging of particles looks like
evaporation of a Bose condensate.

Phenomenological cosmological model (12) we use, describes a transition which
has fixed both beginning and the end ($p = 0$ at the certain moment $t_1$).
The same
result has been obtained using smooth model with $p = ({\gamma(t)-1})
\varepsilon$,
where ${\gamma(t)}\rightarrow 1$ at $t\rightarrow{\infty}$ ${^{15}}$.
Therefore
this result seems to be model-independent.

According to the standard approach (with selfinteraction $\lambda {\phi}^4$,
$\lambda=g_{sc}^2$)
for giving masses to heavy scalars
via the global SSB ${^{1,12,16}}$, we have
  $$M_{sc}\sim  g_{sc} v_I$$
where $v_I$ is a vacuum expectation value which is expected to be
$v_{I}\sim M_{GUT}$.

 The local SSB is expected to give  mass to
the gauge bosons of the order
$$M_x \sim {{g_x v_{II}} }$$
Since both effects occur at similar energy scale $E\sim {E}_{GUT}$,
one can expect that $v_{II} \sim M_{GUT}$.
(Note, that in SUSY case $v_{I} = v_{II} $ and $g_{s} =g_{x}$.)
Hence the equation of state for both scalar and gauge bosons would be the same
up to the constant:
 $$P = const_{(sc,x)} g_{(sc,x)}^{3/2} \rho_{GUT} c^2
\biggl({{E_{GUT}}\over{E_{Pl}}}\biggr)^{5/2}\eqno(25)$$
The estimates presented here
do not depend crucially on the choice of GUT scale as $10^{15}GeV$.
The scale $10^{16}GeV$ can be chosen
(which is preferable by SUSY models).
It is also possible to have both phase transitions
at the same time ${^{17}}$.

We can also estimate characteristic relaxation time for
particle-particle (pp) collisions during a transition and
compare it with the duration of a transition $t_1$.
This relaxation time can be determined as
$$\tau^{pp}_{rel}={1\over{n\sigma_{int}v_T}},\eqno(26)$$
where
$v_T$ is an average thermal velocity of particles. According to (7),
$$v_T=c\biggl({{3kT_H}\over{mc^2}}\biggr)=c\biggl({6\over{\pi}}\biggr)^{1/4}
\biggl({{E_{GUT}}\over{E_{Pl}}}\biggr)^{1/2}\sim {0.01 c}\eqno(27)$$
Then
$$\tau^{pp}_{rel}\sim{1\over{n{\alpha}^2{\lambda}^2c}}
\biggl({E_{Pl}\over{E_{GUT}}}
\biggr)^{1/2}\sim{10^{-32}s}\eqno(28)$$
It represents, in particular, estimate of characteristic timescale
for annihilation.
We see that $\tau^{pp}_{rel}\gg{t_1}$. Therefore equilibrium state for
particles cannot be achieved during a transition. It could mean that
appropriate situation would exist at the energy scale $E\sim E_{GUT}$
for generating
baryon asymmetry by the mechanism proposed by Sakharov ${^{3}}$
which implies nonstationary conditions in the absence of local
thermodynamic equilibrium.

Calculating cosmological model
with the equation of state (25) depending on GUT parameters,
one will have a chance to extract some information concerning GUT
physics from observational cosmological data, and compare it with
accelerator's data. It would provide a way toward astronomical testing
of GUT models.
\vskip0.2in
{\bf Acknowledgements}
\vskip0.1in
We are grateful to W. Krolikowski, M. Olechowski and Z. P\l uciennik
for helpful discussions.
This research was supported in part by the Polish Committee for
Scientific Research through Grant 2-1234-91-01.
\vskip0.2in

{\bf References}
\vskip0.1in
\item{1}
E.W.Kolb, M.S.Turner: 1990, {\sl The Early Universe}, Addison-Wesley Publ.Co.,
New York.
\item{2}
M.S.Turner: 1992, {\sl Inflation After COBE}, FERMILAB-Conf.92/313-A.
\item{3}
A.D.Sakharov: 1967, {\sl Sov. Phys. Lett.} {\bf 5}, 24;
1991, {\sl Sov. Phys. Usp.} {\bf 34}, 417.
\item{4}
A.D.Sakharov: 1966, {\sl Sov. Phys. JETP} {\bf 22}, 241.
\item{5}
E.B.Gliner: 1966, {\sl Sov. Phys. JETP} {\bf 22}, 378.
\item{6}
E.B.Gliner: 1970, {\sl Sov Phys. Dokl.} {\bf 15}, 559.
\item{7}
E.B.Gliner, I.G.Dymnikova: 1975, {\sl Sov. Astr. Lett.} {\bf 1}, 93.
\item{8}
A.Guth: 1981, {\sl Phys. Rev.} {\bf D23}, 389.
\item{9}
L.Kofman, A.Linde, A.Starobinsky: 1994, {\sl Reheating After Inflation},
UH-IFA-94/35; SU-ITP-94-13; YITP/U-94-15.
\item{10}
G.W.Gibbons, S.W.Hawking: 1977, {\sl Phys. Rev.} {\bf D15}, 2738.
\item{11}
V.Amaldi {\it et al.}: 1991, {\sl Phys. Lett.} {\bf B260}, 447.
\item{12}
P.D.B.Collins, A.D.Martin, E.J.Squires: 1991, {\sl Particle Physics and
Cosmology}, John Wiley and Sons.
\item{13}
I.G.Dymnikova: 1986, {\sl Sov. Phys. JETP} {\bf 63}, 1111.
\item{14}
L.D.Landau, E.M.Lifshitz: 1975, {\sl Statistical Physics}, Pergamon Press,
Oxford.
\item{15}
S.Capozziello, I.Dymnikova, R.de Ritis, C.Rubano, and P. Scudellaro: 1994,
submitted to Phys. Lett. A.
\item{16}
C.Quigg: 1983, {\sl Gauge Theories of the Strong, Weak, and Electromagnetic
Interactions}, Addison-Wesley Publ. Co; T.-P. Cheng and L.-F. Li: 1984,
{\sl Gauge Theory of Elementary Particle Physics}, Clarendon Press, Oxford.
\item{17}
I. Dymnikova, M. Krawczyk: 1994,  Equation of state for particles arising
at the Universe at Grand Unification Energies, IFT/9/1994 (hep-ph/9405333);
in {\sl Birth of the Universe}, ed. F. Occhionero, Springer.
\vfill\eject

\end